\documentclass[12pt]{article}
\usepackage{graphicx}
\usepackage{latexsym}
\usepackage{amsmath,amsfonts,amsthm}

\newcommand{\PGf}{\ensuremath{\mathcal{P}}}
\newcommand{\RGf}{\ensuremath{\mathcal{R}}}
\newcommand{\ave}[1]{\langle #1 \rangle}

\begin{document}
\title{A parallel algorithm for the enumeration of self-avoiding polygons 
on the square lattice}
\author{Iwan Jensen \\
Department of Mathematics \& Statistics, The University of Melbourne\\
Vic. 3010, Australia}

\maketitle

\begin{abstract}
We have developed a parallel algorithm that allows us to enumerate the number
of self-avoiding polygons on the square lattice to perimeter length 110.
We have also extended the series for the first 10 area-weighted moments and
the radius of gyration to 100. 
Analysis of the resulting series yields very accurate estimates of the 
connective constant $\mu =2.63815853031(3)$ (biased) and the critical exponent
$\alpha =  0.5000001(2)$ (unbiased). In addition we obtain very accurate
estimates for the leading amplitudes confirming to a high degree of accuracy
various predictions for universal amplitude combinations.
\end{abstract}

\section{Introduction}

A self-avoiding polygon (SAP) on a lattice can be defined as a walk along
the edges of the lattice which starts and ends at the origin but has no other 
self-intersections. Alternatively we can define a SAP as a graph whose vertices 
are of degree 0 or 2 and apart for isolated vertices has only a single component. 
The enumeration of self-avoiding polygons on various lattices is an interesting 
combinatorial problem in its own right, and is also of considerable importance 
in the statistical mechanics of lattice models \cite{HughesV1}. When enumerated
by perimeter SAPs can be considered a model for ring polymers and when enumerated
by area they model vesicles \cite{Leibler87,MEF89a,FGW91}.

The basic problems are the calculation of the number $p_n$ of polygons of perimeter
$n$, the number $a_m$ of polygons of area $m$, or more generally the number 
$p_{m,n}$  of polygons of area $m$ and perimeter $n$. Note that on the square
lattice polygons have an {\em even} perimeter and $p_n =0$ for $n$ odd. 
Here we are interested in area-weighted moments, where the 
$k$'th area-weighted moment is $\ave{a^k}_n = (\sum_m m^k p_{m,n})/p_n$. Also of 
great interest is the mean-square radius of gyration $\ave{R^2}_n$, which measures 
the typical size of a polygon with perimeter $n$. These quantities are expected to 
behave as

\begin{eqnarray}\label{eq:coefgrowth}
p_n & = & B\mu^n n^{\alpha-3}[1+o(1)], \nonumber \\
\ave{a^k}_n & = & E^{(k)}n^{2k\nu}[1+o(1)], \\
\ave{R^2}_n & = & D n^{2\nu}[1+o(1)], \nonumber
\end{eqnarray}
\noindent
where $\mu$ is the so-called connective constant, while $\alpha$ and
$\nu$ are critical exponents. When analyzing the data it is often convenient
to use the associated generating functions

\begin{eqnarray}\label{eq:genfunc}
\RGf^2_g (u)& = &\sum_n n^2 p_{n}\ave{R^2}_n  u^n =
    \sum_{n} r_n u^n \sim R(u)(1-u\mu)^{-(\alpha+2\nu)}, \\
\PGf^{(k)} (u)& = &\sum_n p_{n}\ave{a^k}_n u^n =
    \sum_{n} a^{(k)}_n u^n \sim A^{(k)}(u)(1-u\mu)^{2-(\alpha+2k\nu)}.
\end{eqnarray}
where the various factors are chosen so that $r_n$ and $a^{(k)}_n$ are integers.
These series are thus expected to have a singularity at the critical
point $u_c = 1/\mu$ with critical exponents as above. In particular
we note that the critical exponent of the perimeter generating function,
$\PGf (u)=\PGf^{(0)} (u)$, is $2-\alpha$.

Despite strenuous effort over the past 50 years  or so this problem has not been 
solved on any regular two dimensional lattice.
However, for the hexagonal lattice the critical point, $u_c^2=1/(2+\sqrt{2})$ 
as well as the critical exponents $\alpha = 1/2$ and $\nu =3/4$ are known exactly 
\cite{Nienhuis82a}, though non-rigorously. Very firm evidence exists from
previous numerical work that the exponent $\alpha$ is universal and thus 
equals 1/2 for all two dimensional lattices \cite{GE88a,EG92,JG98,Lin99b}. 
The value of $\nu$ and its universality  have also been confirmed
by numerical work \cite{PR85,GE88a,EG90a,Lin99b,IJ00a}.

It is also known \cite{CG93}
that the amplitude combination $E^{(1)}/D$ is universal, and that
\begin{equation}\label{eq:BDampl}
BD = \frac{5}{32 \pi ^2}\sigma a_0,
\end{equation}
\noindent
where $a_0$ is the area per site and $\sigma$ is an integer such that
$p_n$ is non-zero only if $n$ is divisible by $\sigma$. For the
square lattice $a_0=1$ and $\sigma=2$. These predictions have
been confirmed numerically \cite{CG93,Lin99a,Lin99b,IJ00a,Lin00}.

Recently, Richard {\em et al.} \cite{RGJ01} found, subject to a very reasonable 
conjecture, the exact scaling function for self-avoiding polygons. This in turn 
led to the derivation of universal amplitude combinations for {\em all} the
$E^{(k)}$, namely that $E^{(k)}B^{k-1}$ are known universal constants.
In particular it has been shown that  $E^{(1)} = 1/4\pi$ \cite{JLC94a}.
These predictions were strongly supported by numerical evidence \cite{RGJ01}.

Some years ago \cite{CEG93} it was pointed out that since the hexagonal 
lattice connective constant is given by the zero of a quadratic in $u^2,$ it 
is plausible that this might be the case also for the square lattice 
connective constant.  It was found that $581u^4 + 7u^2 -13$
was the only polynomial  with ``small'' integer coefficients consistent with 
this estimate. The relevant zero of this polynomial is $u_c^2=0.1436806292698685\ldots$.
In \cite{JG99} the numerical evidence was in complete agreement with this conjecture, but 
with 4 more significant digits than when the original suggestion was made. 

This paper builds on the work of Enting \cite{IGE80e} who enumerated square lattice 
polygons to 38 steps using the finite lattice method. Using the same technique this 
enumeration was extended by Enting and Guttmann to 46 steps \cite{EG85} and later 
to 56 steps \cite{GE88a} and further extended to 70 steps
in unpublished work. These extensions to the enumeration were largely
made possible by improved computer technology. Jensen and Guttmann \cite{JG99} 
improved the algorithm and extended the enumeration to 90 steps while 
using essentially  the same computational resources used to obtain 
polygons to 70 steps. The work by Guttmann and Enting \cite{GE88a} also included
calculations of moments of the caliper size distribution. Hiley and Sykes
\cite{HS61} obtained the number of square lattice polygons by both area and
perimeter up to perimeter 18. Enting and Guttmann extended the calculation to
perimeter 42 \cite{EG90a}. The radius of gyration was calculated for SAPs up
to 28 steps by Privman and Rudnick \cite{PR85}. Jensen \cite{IJ00a}
extended the series for area-moments with $k\leq 2$ and the radius of gyration
to 82 steps. In \cite{RGJ01} the calculation for area-moments was extended
to $k \leq 10$.

The main purpose of this paper is to report on a new parallel version of
our earlier algorithms which allows us to significantly extend the series
and use these extended series to critically examine the theoretical predictions
given above as well as revisit the conjecture for the connective constant
on the square lattice. Using the parallelised algorithm
and a new superior memory management, inspired by Knuth's work on the enumeration
of polyominoes \cite{Knuth01}, we have been able to extend the enumeration
of square lattice polygons to 110 steps. We extend the series for area-weighted
moments with $k \leq 10$  and the radius of gyration to 100 steps.

In the next section we will very briefly review the finite lattice
method for enumerating square lattice polygons and give some
details of the improved parallel algorithm. The results of
the analysis of the series are presented in Section~\ref{sec:analysis}
including a detailed discussion of the conjecture for the exact
critical point and numerical tests of the predictions for universal
amplitude combinations.

\section{Enumeration of polygons \label{sec:flm}}

The algorithm used to enumerate SAPs on the square lattice is an enhancement of 
the finite-lattice method devised by Enting \cite{IGE80e} in his pioneering work,
which contains a detailed description of the original approach. A major 
enhancement, resulting in an exponentially more efficient algorithm, is
described in some detail in \cite{JG99} while details of the changes required to 
enumerate area-moments and the radius of gyration can be found in \cite{IJ00a}.
In the following we shall briefly outline those parts of the method required
to understand how the parallel version works. 

\subsection{The transfer matrix algorithm}

The first terms in the series for the polygon generating function can be 
calculated using transfer matrix techniques to count the number of polygons 
in rectangles $W$ vertices wide and $L$ vertices long. Due to the symmetry of 
the square lattice one need only consider rectangles with $L \geq W$. In the 
original application \cite{IGE80e} valid polygons were required to span the 
enclosing rectangle in the lengthwise direction. Clearly polygons which
are narrower than the width of the rectangle are counted many times.
It is, however, easy to obtain the polygons of width exactly $W$ and
length exactly $L$ from this enumeration \cite{IGE80e}. Any polygon spanning such a 
rectangle has a perimeter of length at least $2(W+L)-4$.  By adding the contributions 
from all rectangles of width $W \leq W_{\rm max}$ (where the choice of 
$W_{\rm max}$  depends on available computational resources) and length 
$W \leq L \leq 2W_{\rm max}-W+1$, with contributions from 
rectangles with $L>W$ counted twice, the number of polygons per vertex of an 
infinite lattice is obtained correctly up to perimeter 
$N_{\rm max}=4W_{\rm max}-2$.

\begin{figure}
\begin{center}
\includegraphics{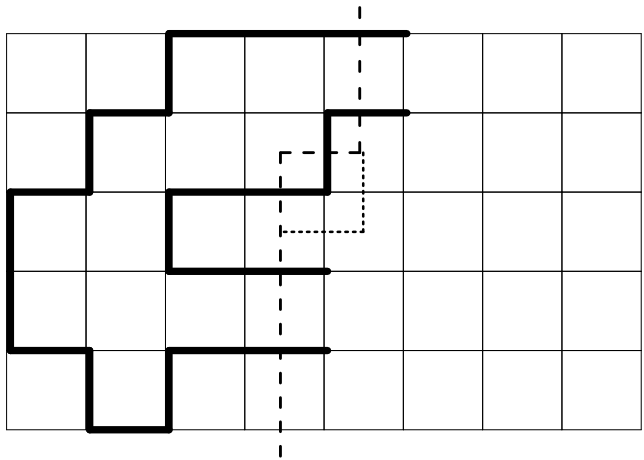}
\end{center}
\caption{\label{fig:transfer}
A snapshot of the boundary line (dashed line) during the transfer matrix 
calculation on the square lattice. Polygons are enumerated by successive
moves of the kink in the boundary line, as exemplified by the position given 
by the dotted line, so that one vertex at a time is added to the rectangle. 
To the left of the boundary line we have drawn an example of a 
partially completed polygon.}
\end{figure}

The transfer matrix technique  involves drawing a boundary line through the
rectangle intersecting a set of up to $W+1$ edges. Polygons in a given rectangle 
are enumerated by moving the boundary line so as to add one vertex at a time, 
as shown in Fig.~\ref{fig:transfer}. In this fashion we build up the rectangle
column by column with each column built up vertex by vertex.  As we move the 
boundary line it intersects partially completed polygons consisting of disjoint 
loops that must all be connected to form a single polygon. For each configuration 
of occupied or empty edges along the intersection we maintain a (perimeter) 
generating function for open loops to the left of the line cutting the 
intersection in that particular pattern. The updating of the generating
functions depends primarily on the configuration of the two edges at the kink
in the boundary line prior to the move (we shall refer to these edges
as the kink edges). As the boundary line is moved the two new edges intersected 
by the boundary line can be either empty or occupied.  

To avoid situations leading to graphs with more than a single
component we have to forbid a loop to close on itself if the boundary line 
intersects any other loops. So two loop ends can only be joined if they belong 
to different loops or all other edges are empty. To exclude loops which close
on themselves we need to label the occupied edges in such a way that
we can easily determine whether or not two loop ends belong to the same loop.
The most obvious choice would be to give each loop a unique label.
However, on two-dimensional lattices there is a more compact scheme
relying on the fact that two loops can never intertwine. Each end of a loop 
is assigned one of two labels depending on whether it is the lower end or 
the upper end of a loop. Each configuration along the boundary line can thus 
be represented by a set of edge states $\{\sigma_i\}$, where

\begin{equation}\label{eq:sapstates}
\sigma_i  = \left\{ \begin{array}{rl}
 0 &\;\;\; \mbox{empty edge},  \\ 
 1 &\;\;\; \mbox{lower end of a loop}, \\
 2 &\;\;\; \mbox{upper end of a loop}. \\
\end{array} \right.
\end{equation}
\noindent
Configurations are read from the bottom to the top. The configuration along the 
intersection of the partially completed polygon in Fig.~\ref{fig:transfer} 
is $\{011\overline{21}22\}$ before the move, where we use over-lining to indicate 
the kink edges, and $\{01\overline{10}022\}$ after the move. 
It is easy to see that this encoding uniquely describes 
which loop-ends are connected. In order to find the upper loop-end, matching a 
given lower end, we start at the lower end and work upwards  in the configuration 
counting the number of `1's and `2's we pass (the `1' of the initial lower end is 
{\em not} included in the count). We stop when the number of `2's  exceeds the 
number of  `1's. This `2'  marks the matching upper end of the loop.  Ignoring the 
`0's the '1's and `2's can be viewed as perfectly balanced parenthesis. Those familiar 
with algebraic languages will recognize that each configuration of labeled
loop-ends forms a Motzkin word \cite{DV84a}. It is known that the number of Motzkin words
of length $m$ grows exponentially like $3^m$. This means that the number of configurations
and thus the computational complexity of the FLM calculation grows like 
$3^{N_{\rm max}/4}$.

\begin{figure}
\begin{center}
\includegraphics[scale=0.5]{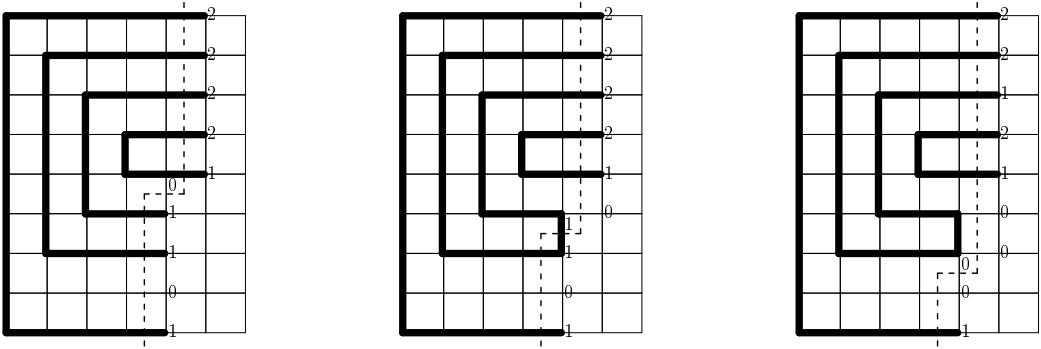}
\end{center}
\caption{\label{fig:joinloops}
Snapshots of the boundary line (dashed line) during the TM calculation. 
Shown is a situation with four nested loops (left panel) where the lower ends
of two loops are joined (middle panel) resulting in a situation with three
nested loops (right panel) and a relabeling of the loop ends.}
\end{figure}

The rules for updating the partial generating functions as the intersection is moved 
are identical to the original work, so we refer the interested reader to \cite{IGE80e} 
for further details regarding this aspect of the transfer matrix calculation.
The only important aspect we wish to emphasize here is that when joining two loop-ends
at the kink we may have to change the labeling of a corresponding loop-end elsewhere
in the resulting new configuration. An example is shown in Fig.~\ref{fig:joinloops}.
In this case we start out with four nested loops corresponding to the
configuration $\{101\overline{10}12222\}$, then upon moving the kink in the
boundary line the lower loop-ends of the second and third loops are joined leading 
to the configuration $\{10\overline{11}012222\}$. After the next move
we see that there are now three differently nested loops and the upper end of the 
second loop (prior to the moves) have become the lower end of the third loop 
(after the moves) resulting in the final configuration $\{1\overline{00}0012122\}$.

The major improvement to the original method as explained in \cite{JG99}
is that we require valid polygons to span the rectangle in 
{\em both} directions. In other words we directly enumerate polygons of
width exactly $W$ and length $L$ rather than polygons of width $\leq W$
and length $L$ as was done originally. At first glance this would appear 
to be inefficient since for many boundary line configurations we 
now have to keep 4 distinct generating functions depending on which borders have been 
touched. However, as demonstrated in practice \cite{JG99} it actually 
leads to an algorithm which is both exponentially faster and whose memory 
requirement is exponentially smaller. Experimentally it was found that
the computational complexity was close to $2^{N_{\rm max}/4}$, much better
than the $3^{N_{\rm max}/4}$ of the original approach.
Realizing the full savings in time and memory usage require
enhancements to the original algorithm. The most important is what
we call {\em pruning}. This procedure, details of which are given
in \cite{JG99}, allows us to discard most of the possible
configurations for large $W$ because they only contribute to polygons of 
length greater than $N_{\rm max}$. Briefly this works as follows.
Firstly, for each configuration we keep track of the current minimum 
number of steps $N_{\rm cur}$ already inserted to the left of the boundary 
line in order to build up that particular configuration. Secondly, we 
calculate the minimum number of additional steps $N_{\rm add}$ required to 
produce a valid polygon. There are three contributions, namely the number 
of steps required to close the polygon, the number of steps needed (if any) 
to ensure that the polygon touches both the lower and upper border, and 
finally the number of steps needed (if any) to extend at least $W$ edges 
in the length-wise direction (remember we only need rectangles
with $L \geq W$). If the sum $N_{\rm cur}+N_{\rm add} > N_{\rm max}$ we 
can discard the partial generating function for that configuration,
and of course the configuration itself, because it won't make a 
contribution to the polygon count up to the perimeter lengths we are 
trying to obtain.

Inspired by Knuth's algorithm for the enumeration of polyominoes \cite{Knuth01},
we implemented a couple of further enhancements to our SAP algorithm.
The first improvement is a superior memory management. A given boundary line
configuration does not contribute until order $N=N_{\rm cur}+N_{\rm add}$,
so we need only retain the first $(N_{\rm max}-N)/2$ terms in the associated 
generating function, the factor of 2 arising since every other term is identically 0. 
In our case the maximum in memory consumption occur at $W=24$where  there are 
approximately $8.1\times 10^8$ distinct configurations and a total of about 
$2.1\times 10^9$ non-zero terms in the generating functions. So on average there is 
only about 2.5 non-zero terms per configuration. The second improvement uses a 
further symmetry of the square lattice. When a column has been completed the 
configuration are symmetric under reflection. That is the generating functions
for the configurations such as, $\{010122000\}$ and $\{000112020\}$, are identical.
This symmetry also extends to the touching of the upper/lower borders of the
rectangle.

The generalization of the algorithm to calculations of area-weighted moments
and the radius of gyration is described in \cite{IJ00a}. Note that the
additional symmetry mentioned above does not extend to the radius of gyration
calculation.

\subsection{Parallelization}
\label{SECpara}

The computational complexity of the FLM grows exponentially with
the number of terms one wishes to calculate. It is therefore 
little wonder that implementations of the algorithms have always
been geared towards using the most powerful computers available. In the
past decade or so parallel computing has become the paradigm
for high performance computing. The early machines were largely dedicated
MPP machines which more recently have been super-seeded by clusters.

The transfer-matrix algorithms used in the calculations of the
finite lattice contributions are eminently suited for parallel
computations. 

The most basic concerns in any efficient parallel algorithm is
to minimise the communication between processors and ensure that
each processor does the same amount of work and use the same amount 
of memory. In practice one naturally has to strike some compromise
and accept a certain degree of variation across the processors.

One of the main ways of achieving a good parallel algorithm using 
data decomposition is to try to find an invariant under the
operation of the updating rules. That is we seek to find some property
about the configurations along the boundary line which
does not alter in a single iteration.
The algorithm for the enumeration of polygons is quite complicated 
since not all possible configurations occur due to pruning
and an update at a given set of edges might change the state of 
an edge far removed, e.g., when two lower loop-ends are joined
we have to relabel one of the associated upper loop-ends as
a lower loop-end in the new configuration (see Fig.~\ref{fig:joinloops}).
However, there still is an invariant since any edge not
directly involved in the update cannot change from being 
empty to being occupied and vice versa. That is only the kink edges 
can change their occupation status. This invariant
allows us to parallelise the algorithm in such a way
that we can do the calculation completely independently on each
processor with just two redistributions of the 
data set each time an extra column is added to the lattice. 

The main points of the algorithm are summarized below:

\begin{enumerate}
\item With the boundary line in an upright position distribute the
data across processors so that configurations with the same 
occupation pattern along the {\em lower} half of the boundary line 
are placed on the same processor. 
\item Do the TM update inserting the top-half of a new column.
This can be done {\em independently} by each processor because the 
occupation pattern in the lower half remains unchanged.
\item Upon reaching the half-way mark redistribute the data
so that configurations with the the same occupation pattern along 
the {\em upper} half of the boundary line are placed on the same processor. 
\item Do the TM update inserting the bottom-half of a new column.
\item Go back to 1.
\end{enumerate}

The redistribution among processors is done as follows: 

\begin{enumerate}
\item On each processor run through the configurations to establish
the configuration pattern $c$ of each configuration and calculate, $n(c)$, the
number of configurations with a given pattern.
\item Calculate the sum of $n(c)$ on say processor 0.
\item Sort $n(c)$ on processor 0.
\item On processor 0 assign each pattern to a processor $p(c)$ such that:
\begin{enumerate} 
\item Set $p_{id}=0$.
\item Assign the {\em most} frequent unassigned pattern $c$ to processor $p_{id}$. 
\item If the number of configurations assigned to $p_{id}$ is
less than the number of configurations assigned to $p_{0}$ then
assign the {\em least} frequent  unassigned patterns to $p_{id}$ until
the desired inequality is achieved.
\item set $p_{id} = p_{id} \mod N_p$, where $N_p$ is the number of processors.
\item Repeat from (b) until all patterns have been assigned.
\end{enumerate}
\item Send $p(c)$ to all processors.
\item On each processor run through the configurations sending
each configuration to its assigned processor. 
\end{enumerate}

\begin{table}
\caption{ \label{tab:para}
CPU-time and memory use for the parallel algorithm for
enumerating polygons of maximal perimeter 98 at width 22.}
\begin{center}
\begin{tabular}{rrrrrrr}
\hline
Proc. & CPU time & Elapsed time & Max Conf & Min Conf & Max Term & Min Term \\
\hline
1     &  33:26   &    33:34:30  & 94858092 &          & 202124719 & \\
2     &  34:58   &    17:31:09  & 45332715 & 45312242 & 99729074 & 99050619 \\
4     &  34:15   &     8:35:57  & 22762665 & 22667218 & 51880015 & 51263646 \\
8     &  34:03   &     4:16:51  & 11692292 & 11525456 & 26498730 & 26097260 \\
16    &  34:16   &     2:09:40  &  5880705 &  5707628 & 13523912 & 13037482 \\
32    &  33:15   &     1:03:04  &  2941787 &  2821055 &  6934653 &  6451282 \\
64    &  32:29   &       31:07  &  1489116 &  1398768 &  3519013 &  3222199 \\
\hline
\end{tabular}
\end{center}
\end{table}

The bulk of the calculations were performed on the facilities of the Australian
Partnership for Advanced Computing (APAC). The APAC facility is a Compaq Server 
Cluster with 125 ES45's each with 4 1 Ghz chips for a total of 500 processors in 
the compute partition. The cluster has a total peak speed over 1Tflop. Each server 
node has at least 4 Gb of memory. 
Nodes are interconnected by a fat-tree low latency (MPI $<$ 5 usecs), 
high bandwidth (250 Mb/sec bidirectional) Quadrics network. 

In Table~\ref{tab:para} we have listed the time and memory use of the
algorithm for $N_{\rm max} = 98$ at $W=22$ using from 1 to 64 processors.
The memory use of the single processor job was about 3Gb. As can be seen
the algorithm scales perfectly from 1 to 64 processors since the total
CPU time (column 2) stays almost constant while the elapsed time is
halved when the number of processors is doubled. We expect that the 
rather surprising drop in CPU time at 32 or 64 processors is caused by
better single processor optimization by the compiler. One would for example
expect that the average time taken to fetch elements from main memory drops
as the memory size on each individual processor drops from 3Gb for the 
computation using a single processor to just under 50Mb for the 64 processor
computation. Another main issue in parallel computing is that of load 
balancing, that is, we wish to ensure to the greatest extent possible that
the workload is shared equally among all the processors. As can be seen
this algorithm is quite well balanced. Even with 64 processors, where each
processor uses only about 50Mb of memory, the difference between the processor
handling the maximal and minimal number of configurations is less than 10\%.
The same holds true for the total number of terms retained in the generating
functions. 

A simple timing of the various sub-routines of the parallel algorithm shows 
that the typical time to do a redistribution is the same as the average time 
taken per iteration in order to move the kink once. Since the maximal time
use at $N_{\rm max}=110$ occurs at $W=24$ there are 24 iteration and just 
2 redistributions per added column, so the overall cost of parallel execution 
is smaller than 10\%.

\subsection{Further details}

Finally a few remarks of a more technical nature. The number of contributing 
configurations becomes very sparse in the total set of possible states along 
the boundary line and as is standard in such cases one uses a hash-addressing 
scheme. Since the integer coefficients occurring in the series 
expansion become very large, the calculation was performed using modular 
arithmetic \cite{KnuthACPv2}. This involves performing the calculation modulo 
various integers $p_i$ and then reconstructing the full integer
coefficients at the end. The $p_i$ are called moduli and must be chosen
so they are mutually prime, e.g., none of the $p_i$ have a common divisor.
The Chinese remainder theorem ensures that any integer has a unique 
representation in terms of residues. If the largest absolute values occurring 
in the final expansion is $m$, then we have to use a number of moduli $k$ 
such that $p_1p_2\cdots p_k/2 > m$. Since we are using a heavily loaded 
shared facility  CPU time was more of a immediate limitation than memory 
and secondly more memory was used for the date required to specify the 
configuration and manage the storage than for storing the actual terms of 
the generating functions. So we used the moduli $p_0=2^{62}$, $p_1=2^{62}-1$ 
and $p_2=2^{62}-3$, which allowed us to represent $p_n$ correctly using these 
three moduli. 
The calculation of the area-weighted moments and the radius of gyration 
require a lot more memory for the generating functions (plus the radius
of gyration calculation involves multiplication with quite large integers)
so in this case we used prime numbers of the form $2^{30}-r_i$ for the
moduli $p_i$. Up to 6 primes were needed to represent the coefficients
correctly. 

Combining all the memory minimization tricks mentioned above allows us to 
extend the series for the square lattice polygon generating function from 
90 terms to 110 terms using at most 36Gb of memory. The calculations
requiring the most resource were at widths 23--25. These cases were done
using 40 processors and took about 8-10 hours each per prime.
The total CPU time required was about 1500 hours per prime. Obviously the 
calculation for each width and prime are totally independent and several 
calculations can be done simultaneously. A similar total amount of resources
was required to calculate the area-moments and the radius of gyration.

In Table~\ref{tab:series} we have listed the new terms obtained
in this work for the number of polygons with perimeter 92--110. The number of polygons 
of length $\leq 56$ can be found in \cite{GE88a} while those up to length 90 are 
listed in \cite{JG99}.

\begin{table}
\caption{\label{tab:series} The number, $p_n$, of embeddings of 
$n$-step polygons on the square lattice. Only non-zero terms are listed.}
\begin{center}
\scriptsize
\begin{tabular}{rrrr} \hline \hline
$n$ & $p_n$ & $n$ & $p_n$ \\ \hline 
92  & 3959306049439766117380237943449096   
    &  102  & 49985425311177130573540712929060556804 \\
94  & 26117050944268596220897591868398452 
    &  104  & 331440783010043009106782321492277936522 \\
96  & 172472018113289556124895798382016316 
    &  106  &  2199725502650970871182263620080571090156 \\
98  & 1140203722938033441542255979068861816 
    &  108  &  14612216410979678692651320184958285074180 \\
100 & 7545649677448506970646886033356862162   
    &  110  & 97148177367657853074723038687712338567772 \\
\hline \hline
\end{tabular}
\end{center}
\end{table}

\section{Analysis of the series \label{sec:analysis}}

To obtain the singularity
structure of the generating functions we used the numerical method of
differential approximants \cite{AJG89a}. Since all odd terms in the series
are zero and the first non-zero term is $p_4$ we actually analyzed the
function $P(u)=\sum_n p_{2n+4} u^n$, and so on. These functions have critical 
points at $u=u_c^2$ with the same exponents as those of (\ref{eq:genfunc}). 
Our main objective is to obtain very accurate estimates for the connective 
constant $\mu$ and the critical exponents $\alpha$ and $\nu$. In particular we are keen
to test a conjecture \cite{CEG93} for the exact value of the connective constant 
and confirm to a very high degree of precision the exact values of
the exponents.

Once the exact values of the exponents have been confirmed we will turn our
attention to the ``fine structure'' of the asymptotic form of the
coefficients. In particular we are interested in obtaining accurate
estimates for the amplitudes $B$, $D$ and $E^{(k)}$. We do this
by fitting the coefficients to the assumed form (\ref{eq:coefgrowth}).
In this case our main aim is to test the validity of the predictions
for the amplitude combinations mentioned in the Introduction.

\subsection{The polygon generating function}

In Table~\ref{tab:analysis}  we have listed estimates for the
critical point $u_c^2$  and exponent $2-\alpha$
of the series for the square lattice SAP generating function.  
The estimates were obtained by averaging values obtained from second and
third order differential approximants. For each order $L$ of the inhomogeneous 
polynomial we averaged over those approximants to the series which used  
at least the first 45 terms of the series (that is, polygons of perimeter at least 90).
The error quoted for these estimates reflects the spread (basically one standard
deviation) among the approximants. Note that these error bounds should
{\em not} be viewed as a measure of the true error as they cannot include
possible systematic sources of error. Based on these estimates we 
conclude that $u_c^2 = 0.14368062925(5)$ and $\alpha = 0.5000001(2)$.
This analysis adds strongly to the already very convincing evidence that 
the critical exponent $\alpha = 1/2$ exactly.

\begin{table}[h]
\caption{\label{tab:analysis} Estimates for the critical point
$u_c^2$ and exponent $2-\alpha$ obtained from second and third order
differential approximants to the series for square lattice
polygon generating function. $L$ is the order of the inhomogeneous
polynomial.}
\begin{center}
\begin{tabular}{lllll} \hline \hline
 $L$   &  \multicolumn{2}{c}{Second order DA} & 
       \multicolumn{2}{c}{Third order DA} \\ \hline 
    &  \multicolumn{1}{c}{$u_c^2$} & \multicolumn{1}{c}{$2-\alpha$} & 
      \multicolumn{1}{c}{$u_c^2$} & \multicolumn{1}{c}{$2-\alpha$} \\ \hline
0    & 0.143680629242(28) & 1.500000116(94) &
       0.143680629246(22) & 1.500000105(73) \\
2    & 0.143680629245(15) & 1.500000111(63) &
       0.143680629247(21) & 1.500000097(81) \\
4    & 0.143680629246(16) & 1.500000107(62) &
       0.143680629251(22) & 1.500000080(99) \\
6    & 0.143680629250(17) & 1.500000094(65) &
       0.143680629244(22) & 1.500000109(72) \\
8    & 0.143680629249(22) & 1.500000094(72) &
       0.143680629249(28) & 1.50000009(14) \\
10   & 0.143680629248(19) & 1.500000095(66) &
       0.143680629252(28) & 1.50000006(15) \\
12   & 0.143680629246(21) & 1.500000105(70) &
       0.143680629247(18) & 1.500000100(70) \\
14   & 0.143680629242(20) & 1.500000116(66) &
       0.143680629245(26) & 1.500000099(99) \\
16   & 0.143680629252(18) & 1.500000086(63) &
       0.143680629247(25) & 1.500000097(94) \\
18   & 0.143680629254(15) & 1.500000076(65) &
       0.143680629247(22) & 1.500000098(81) \\
20   & 0.143680629238(26) & 1.500000122(74) &
       0.143680629242(23) & 1.500000113(87) \\
\hline \hline
\end{tabular}
\end{center}
\end{table}

\begin{figure}
\begin{center}
\includegraphics{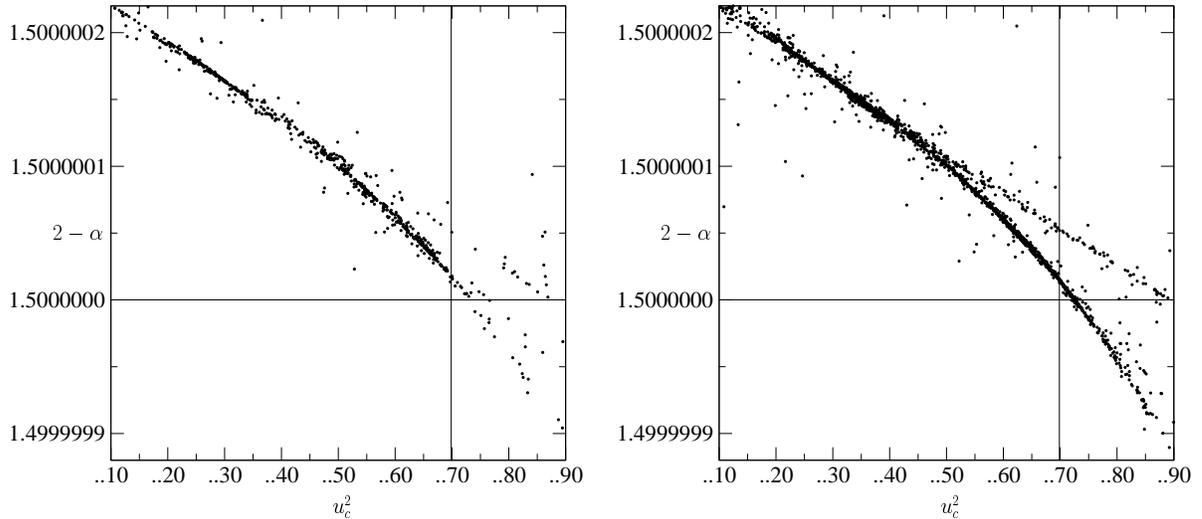}
\end{center}
\caption{\label{fig:crpexp} Estimates for the critical exponent
$2-\alpha$ vs. estimates for the critical point $u_c^2$ of the
square lattice polygon generating function. Each tick label along the
$x$-axis is preceded by the value 0.1436806292. The straight lines
correspond to $2-\alpha=3/2$ and $u_c^2=0.1436806292698685\ldots$.}
\end{figure}

If we take the conjecture $\alpha =1/2$ to be true we can obtain a refined estimate 
for the critical point $u_c^2$ enabling us to check whether or not
the estimates for $u_c^2$ still agree with the root of the polynomial.
In Fig.~\ref{fig:crpexp} we have plotted estimates
for the critical exponent $2-\alpha$ against estimates for the 
critical point $u_c^2$. Each dot in the left (right) panel of this 
figure represents a pair of estimates obtained from a second (third) order 
inhomogeneous differential approximant. The order of the inhomogeneous 
polynomial was varied from 0 to 10. As can be seen the estimates for
the critical exponent cross the line $2-\alpha = 3/2$ at a value 
$u_c^2 \simeq 0.143680629273$, which is slightly larger than the value 
obtained from the root of the polynomial suggested as possibly providing the
exact value. So this is the first direct evidence that the 
conjecture could be wrong. Since the difference only occurs
in the 12th significant digit we do not feel confident that the numerical
evidence alone is sufficient to disprove the conjecture. It may well
be the case that there are subtle systematic trends in the estimates, which
preclude them from having converged to the true values of the parameters.
However, as emphasized in \cite{JG99} the other zero of the polynomial is at 
$u_c^2 = -0.1557288\ldots,$ and as was the case in this previous analysis, we see 
no evidence of such a singularity, which casts serious doubt on the validity of 
the conjecture. Particularly since we are not aware of any arguments as to why  
we might not expect to see the  singularity on the negative real axis from our 
series analysis. Taken together these two pieces of `evidence' may well
be sufficient disprove to the conjecture. Ultimately we will let the
reader make their own judgment. 

Based on this analysis we adopt the value $u_c^2 = 0.143680629273(3)$
and thus $\mu =2.63815853031(3)$ as our final estimates.

\subsection{The radius of gyration and area-weighted moments}

\begin{table}[h]
\caption{\label{tab:rgmomana} Estimates for the critical point
$u_c^2$ and exponents $-(\alpha+2\nu)$ and $2-(\alpha+2\nu)$ obtained 
from second (top half) and third (bottom half) order differential approximants 
to the series for the radius of gyration and first area-weighted moment of 
square lattice SAP}
\begin{center}
\begin{tabular}{lllll} \hline \hline
 Series:   &  \multicolumn{2}{c}{$\RGf^2_g (u)$} & 
       \multicolumn{2}{c}{$\PGf^{(1)}(u)$} \\ \hline 
 $L$   &  \multicolumn{1}{c}{$u_c^2$} & \multicolumn{1}{c}{$2-\alpha$} & 
      \multicolumn{1}{c}{$u_c^2$} & \multicolumn{1}{c}{$2-\alpha$} \\ \hline
0    & 0.1436805865(92) & -1.999681(36)
     & 0.1436806053(50) &  0.000122(15) \\
2    & 0.143680592(10)  & -1.999704(43)
     & 0.143680609(10)  &  0.000143(60) \\
4    & 0.1436805889(82) & -1.999689(35)
     & 0.143680609(11)  &  0.000139(50) \\
6    & 0.143680583(23)  & -1.999676(82)
     & 0.143680604(12)  &  0.00007(16) \\
8    & 0.143680588(10)  & -1.999680(55)
     & 0.143680608(10)  &  0.00010(10) \\
10   & 0.143680591(12)  & -1.999703(59)
     & 0.143680616(22)  & -0.00011(79) \\
\hline
0    & 0.1436806081(85) & -1.999822(53)
     & 0.143680607(12)  &  0.00021(28) \\
2    & 0.143680605(13)  & -1.999803(79)
     & 0.143680616(11)  &  0.00011(12) \\
4    & 0.1436806074(92) & -1.999812(61)
     & 0.1436806143(73) &  0.000108(42) \\
6    & 0.143680606(11)  & -1.999817(71)
     & 0.1436806166(64) &  0.000095(26) \\
8    & 0.1436806057(93) & -1.999809(53)
     & 0.1436806148(45) &  0.000083(40) \\
10   & 0.143680606(11)  & -1.999817(61)
     & 0.1436806154(55) &  0.000099(21) \\
\hline \hline
\end{tabular}
\end{center}
\end{table}

Table~\ref{tab:rgmomana} contains estimates for $u_c^2$ and the critical
exponents of the generating functions (\ref{eq:genfunc}) for the radius
of gyration and first area-weighted moment. Suffice to say, the estimates
of the exponents are in agreement with the conjectured exact value $\nu=3/4$.

\subsection{The amplitudes}

\begin{figure}
\begin{center}
\includegraphics[scale=0.4]{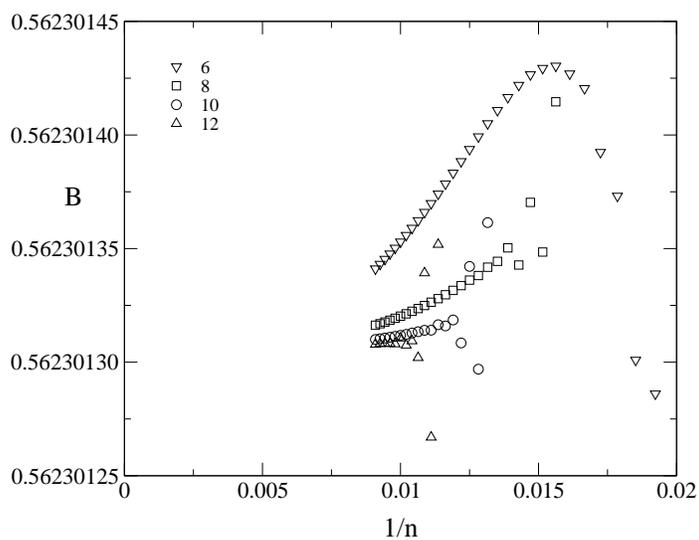}
\end{center}
\caption{\label{fig:sapampl} 
Estimates for the amplitude B vs. $1/n$. Each data set is obtained by fitting 
$p_n$ to the form (\protect{\ref{eq:sapasymp}}) using from 6 to 12 correction terms.}
\end{figure}

The asymptotic form of the coefficients $p_n$ of the polygon generating 
function has been studied in detail previously \cite{CG96,JG99}.
As argued in \cite{CG96} there is no sign of non-analytic 
corrections-to-scaling exponents to the polygon generating function 
and one therefore finds that

\begin{equation}\label{eq:sapasymp}
p_n = \mu^n n^{-5/2} \sum_{i\geq 0} a_i/n^i.
\end{equation}
\noindent
This form was confirmed with great accuracy in \cite{JG99}. 
Estimates for the leading amplitude $B=a_0$ can thus be obtained by
fitting $p_n$ to  the form given in equation~(\ref{eq:sapasymp}) using
increasing values of $k$. As in \cite{IJ00a} we find it useful to
check the behaviour of  the estimates by plotting the results for the 
leading amplitude vs. $1/n$ (see Fig.~\ref{fig:sapampl}), where $p_n$ 
is the last term used in the fitting, and $n$ is varied from 110 down to
50. We again notice that as more and more correction terms are added to the 
fits the estimates exhibits less curvature and that the slope become less steep. 
This is very strong evidence that (\ref{eq:sapasymp})  indeed is the correct 
asymptotic form of $p_n$. We estimate that $B=0.56230130(2)$.

\begin{figure}
\begin{center}
\includegraphics[scale=1]{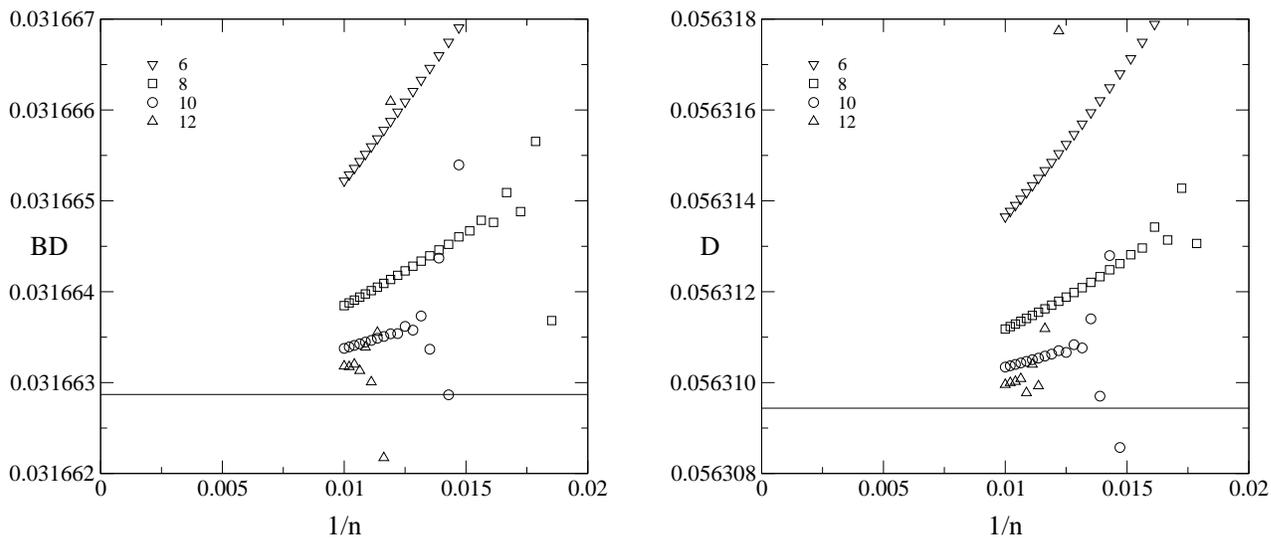}
\end{center}
\caption{\label{fig:rgsampl} 
Estimates for the amplitudes BD  and D vs. $1/n$. Each data set is obtained by fitting 
$r_n$ to the form (\protect{\ref{eq:rgsasymp}}) and $r_n/p_n$ 
to the form (\protect{\ref{eq:rgsratasymp}}) while using from 6 to 12 correction terms.}
\end{figure}

The asymptotic form of the coefficients $r_n$ in the generating function
for the radius of gyration was studied in \cite{IJ00a}. When fitting 
to a form similar to equation~(\ref{eq:sapasymp}), assuming that there are 
only analytic corrections-to-scaling, we found that the amplitudes of higher
order terms are very large and that the leading amplitude converge rather
slowly. This indicates that this asymptotic form is incorrect. We found
that the coefficients fit better if we assume a leading non-analytic 
correction-to-scaling exponent $\Delta=3/2$. This result confirms 
the prediction of Nienhuis \cite{Nienhuis82a}. Note, that since the polygon 
generating function exponent $2-\alpha = 3/2$ a correction-to-scaling 
exponent $\Delta=3/2$ is perfectly consistent with the asymptotic form
(\ref{eq:sapasymp}). Because  $2-\alpha+\Delta$ is an integer the 
non-analytic correction term becomes part of the analytic background
term \cite{CG96}. We thus proposed the following asymptotic form:

\begin{equation}\label{eq:rgsasymp}
r_n = \mu^n n [BD + \sum_{i\geq 0} a_i/n^{i/2}].
\end{equation}
\noindent
Alternative we could fit to the form
\begin{equation}\label{eq:rgsratasymp}
r_n/p_n = n^{7/2} [D + n^{5/2}\sum_{i \geq 0} a_i/n^{i/2}].
\end{equation}
\noindent
In figure~\ref{fig:rgsampl} we show the leading amplitudes resulting from 
such fits while using from 1 to 10 terms in these expansions. Also shown
in these figures (solid lines) are the predicted exact value of $BD$, given
in equation~{\ref{eq:BDampl}, and the prediction for $D$ using the estimate 
for $B$ obtained above. As can be seen the leading amplitudes clearly 
converge towards their expected values and from these plots we can conclude 
that the prediction for $BD$ has been confirmed to at least 6 digit accuracy.
Assuming that  equation~({\ref{eq:BDampl}) is exact and using the very 
accurate estimate for $B$ we find that $D=0.056309437(2)$.

Next we test the predictions \cite{RGJ01} for the amplitudes of the
area-weighted moments. We fit the the coefficients to the assumed form
\begin{equation}\label{eq:momampl}
 a_n^{(k)} \approx \mu^{n} n^{(\alpha+2k\nu)-1}[E^{(k)}+\sum_{i\ge 0}a_i/n^{1+i/2}].
\end{equation}
We obtain several data sets by varying the number of terms using in this formula from
8 to 12. To obtain the final estimates we do a simple linear regression on the data
for the amplitudes as a function of $1/n$ extrapolating to $1/n \rightarrow 0$.
We estimate the error on the amplitude estimate from the spread among the different 
data sets.
In this way, we obtain the results for the leading amplitudes listed 
in Table~\ref{tab:momampl}.

\begin{table}
\caption{\label{tab:momampl}
Exact values and estimates from square lattice polygons for
the universal amplitude combinations.}
\begin{center}
\begin{tabular}{lll}
\hline \hline
Amplitude  & Exact value &  Estimate \\
\hline
$E^{(1)}$    & $0.795774715\ldots \times 10^{-1}$  & $0.795773(2)\times 10^{-1}$ \\
$E^{(2)}B$   & $0.335953483\ldots \times 10^{-2}$  & $0.335952(2)\times 10^{-2}$ \\
$E^{(2)}B^2$ & $0.100253732\ldots \times 10^{-3}$  & $0.100253(1)\times 10^{-3}$ \\
$E^{(2)}B^3$ & $0.237553411\ldots \times 10^{-5}$  & $0.237552(2)\times 10^{-5}$ \\
$E^{(2)}B^4$ & $0.475738345\ldots \times 10^{-7}$  & $0.475736(3)\times 10^{-7}$ \\
$E^{(2)}B^5$ & $0.836630215\ldots \times 10^{-9}$  & $0.836624(5)\times 10^{-9}$ \\
$E^{(2)}B^6$ & $0.132514776\ldots \times 10^{-10}$ & $0.132514(2)\times 10^{-10}$\\
$E^{(2)}B^7$ & $0.192419637\ldots \times 10^{-12}$ & $0.192418(2)\times 10^{-12}$\\
$E^{(2)}B^8$ & $0.259465635\ldots \times 10^{-14}$ & $0.259464(2)\times 10^{-14}$\\
$E^{(2)}B^9$ & $0.328063262\ldots \times 10^{-16}$ & $0.328062(4)\times 10^{-16}$\\
\hline \hline
\end{tabular}
\end{center}
\end{table}

It is clear that the results for the first 10 area weighted moments
are in excellent agreement with the numerical estimates. On
this basis we conclude that the conjectured scaling function and
derived exact amplitude combinations \cite{RGJ01} are correct.

\section{Conclusion}

We have presented an improved and parallel algorithm for the enumeration 
of self-avoiding polygons on the square lattice. This algorithm has enabled 
us to obtain polygons up to perimeter length 110 and their radius of gyration
and area-weighted moments up to perimeter 100. Our extended series enables 
us to give an extremely precise estimate of the connective constant
$\mu =2.63815853031(3)$. This estimate provides the first direct evidence
that  an earlier conjecture for the exact value of $\mu$ could be incorrect.
We confirmed to a very high degree of accuracy
the predicted exponent values $\alpha=1/2$ and $\nu=3/4$. 
We also obtained very accurate estimates for the leading amplitude
$B=0.56230130(2)$ of the asymptotic expansion of $p_n$, and confirmed 
the correctness of theoretical predictions for the values of the 
amplitude combinations $BD$ and $E^{(k)}B^{k-1}$.

\section*{E-mail or WWW retrieval of series}

The series for the generating functions studied in this paper 
can be obtained via e-mail by sending a request to 
I.Jensen@ms.unimelb.edu.au or via the world wide web on the URL
http://www.ms.unimelb.edu.au/\~{ }iwan/ by following the instructions.

\section{Acknowledgments}

The calculations presented in this paper would not have been possible
without a generous grant of computer time on the server cluster of the
Australian Partnership for Advanced Computing (APAC). We also used
the computational resources of the Victorian Partnership for Advanced 
Computing (VPAC). We gratefully acknowledge financial support from 
the Australian Research Council.

\bibliography{sap,series}

\end{document}